\documentclass[fleqn,usenatbib]{mnras}

\usepackage{newtxtext,newtxmath}
\usepackage{chemformula}

\usepackage[T1]{fontenc}
\usepackage[version=4]{mhchem}

\usepackage{longtable}

\DeclareRobustCommand{\VAN}[3]{#2}
\let\VANthebibliography\thebibliography
\def\thebibliography{\DeclareRobustCommand{\VAN}[3]{##3}\VANthebibliography}

\usepackage{graphicx}	% Including figure files
\usepackage{amsmath}	% Advanced maths commands

\title[Laboratory rotational spectroscopy and interstellar search for the protein precursor 4-oxobutanenitrile]
{Laboratory rotational spectroscopy and interstellar search for the protein precursor 4-oxobutanenitrile (HCOCH$_2$CH$_2$CN)}

\author[V. M. Rivilla et al.]{
V. M. Rivilla,$^{1}$
\thanks{vrivilla@cab.inta-csic.es; emiliojose.cocinero@ehu.eus; sonia.melandri@unibo.it }
\thanks{These authors contributed equally to this work.}
E. R. Alonso,$^{2,3,4}$
${\color{blue}\dagger}$
W. Song,$^{5}$
${\color{blue}\dagger}$
A. Insausti,$^{2,3}$
A. Maris,$^{5}$
F. J. Basterretxea,$^{2}$
\newauthor
S. Melandri,$^{5}$
$^{\color{blue}\star}$
I. Jiménez-Serra,$^{1}$
and E. J. Cocinero$^{2,3}$
$^{\color{blue}\star}$
\\
$^{1}$Centro de Astrobiolog\'ia (CAB, CSIC-INTA), Ctra. de Ajalvir Km. 4, Torrej\'on de Ardoz, 28850 Madrid, Spain\\
$^{2}$Departamento de Química Física, Facultad de Ciencia y Tecnología, Universidad del País Vasco (UPV/EHU), Campus de Leioa, Ap. 644, 48080 Bilbao, Spain\\
$^{3}$Instituto Biofisika (CSIC, UPV/EHU), 48080 Bilbao, Spain \\
$^{4}$ {\rm Current address:} Grupo de Espectroscopía Molecular (GEM), Edificio Quifima, Laboratorios de Espectroscopia y Bioespectroscopia, \\ Universidad de Valladolid, 47011 Valladolid, Spain\\
$^{5}$ Department of Chemistry "Giacomo Ciamician", University of Bologna, via P. Gobetti, 85, 40129 Bologna, Italy\\
}

\date{Accepted . Received ; in original form }

\pubyear{2025}

\begin{document}
\label{firstpage}
\pagerange{\pageref{firstpage}--\pageref{lastpage}}
\maketitle

\begin{abstract}
%It should be a single paragraph not more than 250 words (200 words for Letters).

Understanding the presence and distribution of prebiotic precursors in the interstellar medium (ISM) is key to tracing the chemical origins of life. Among them, 4-oxobutanenitrile (\ch{HCOCH2CH2CN}) has been identified in laboratory simulations as a plausible intermediate in the formation of glutamic acid, a proteinogenic amino acid. 
Here, we report its gas-phase rotational spectrum, measured using two complementary techniques: chirped-pulse Fourier transform microwave (CP-FTMW) spectroscopy (2$-$18 GHz) and free-jet millimeter-wave (FJ$-$AMMW) absorption spectroscopy (59.6$-$80 GHz). 
Quantum chemical calculations revealed nine low-energy conformers, of which the TC conformer was assigned based on the measured spectra. 
The resulting spectroscopic parameters were used to search for the molecule in the ultradeep spectral survey of the G+0.693-0.027 molecular cloud, located in the Galactic Center. No signal attributable to 4$-$oxobutanenitrile was detected. A stringent upper limit to its column density was derived ($N<$ 4 $\times$10$^{12}$ cm$^{-2}$), corresponding to a molecular abundance of $<$ 2.9 $\times$10$^{-11}$ relative to H$_{2}$. This upper limit lies well below the observed abundances of simpler structurally related species containing $-$HCO and $-$CN groups, underscoring the challenge of detecting increasingly complex prebiotic molecules in the ISM and the need for future, more sensitive astronomical facilities.

\end{abstract}


%%%%%%%%%%%%%%%%%%%%%%%%%%%%%%%%%%%%%%%%%%%%%%%%%%

%%%%%%%%%%%%%%%%% BODY OF PAPER %%%%%%%%%%%%%%%%%%

\section{Introduction}

\begin{figure}
     \centering
    \includegraphics[width=0.99\linewidth]{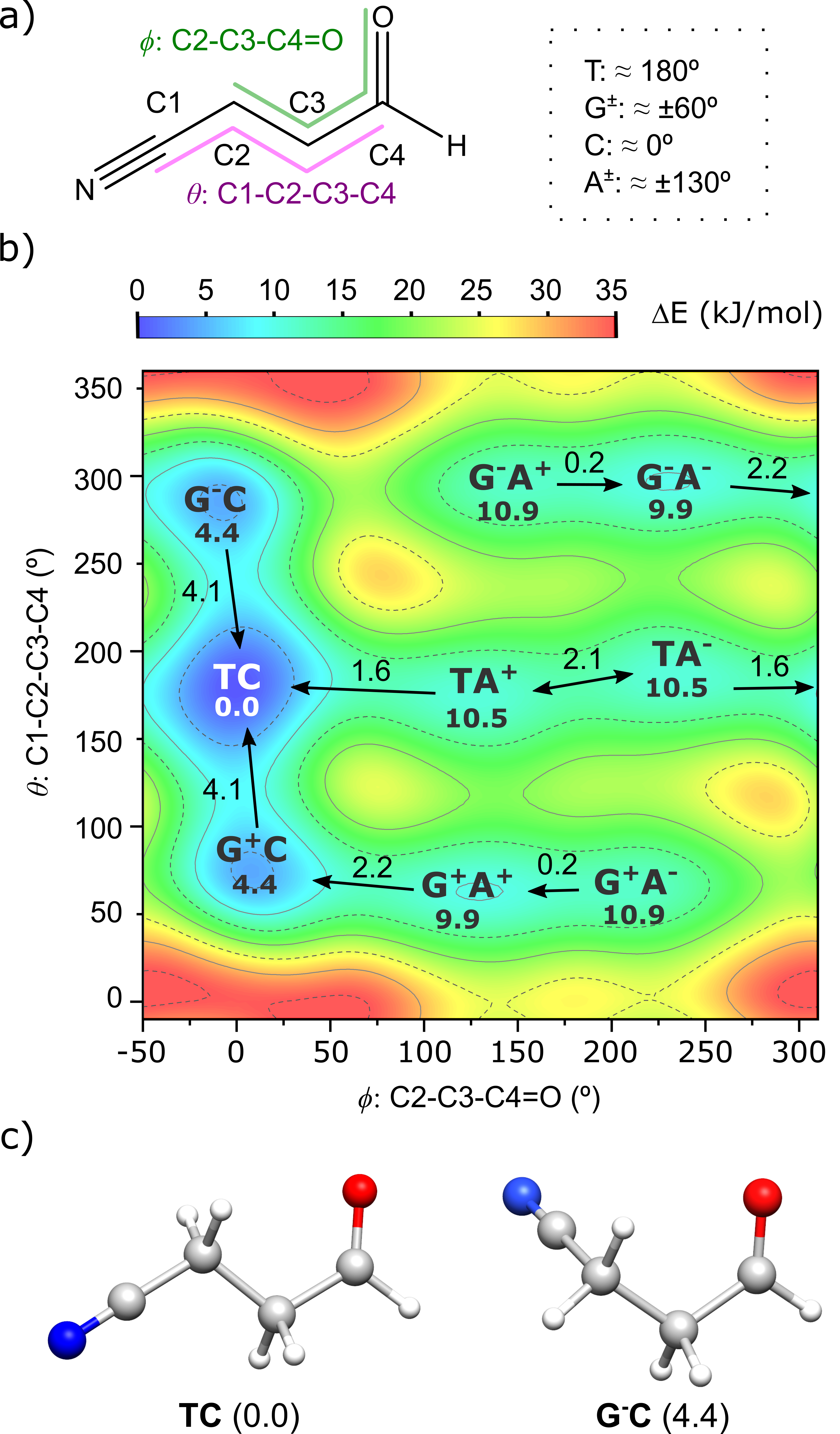}
      \caption{(a) Definition of the torsional angles $\theta$ = C1-C2-C3-C4 and $\phi$ = C2-C3-C4=O used to explore the conformational space of 4-oxobutanenitrile. (b) Relaxed 2D potential energy surface (PES) computed at the B3LYP-D3BJ/def2-TZVPP level showing nine minima. Arrows indicate barriers between key conformers (in kJ mol$^{-1}$). (c) Optimized geometries of the global minimum TC and the second-lowest conformer G$^{-}$C/ G$^{+}$C, with their relative energies including ZPE.}
         \label{fig:conformeros}
\end{figure}

\begin{figure*}
     \centering
    \includegraphics[width=0.99\linewidth]{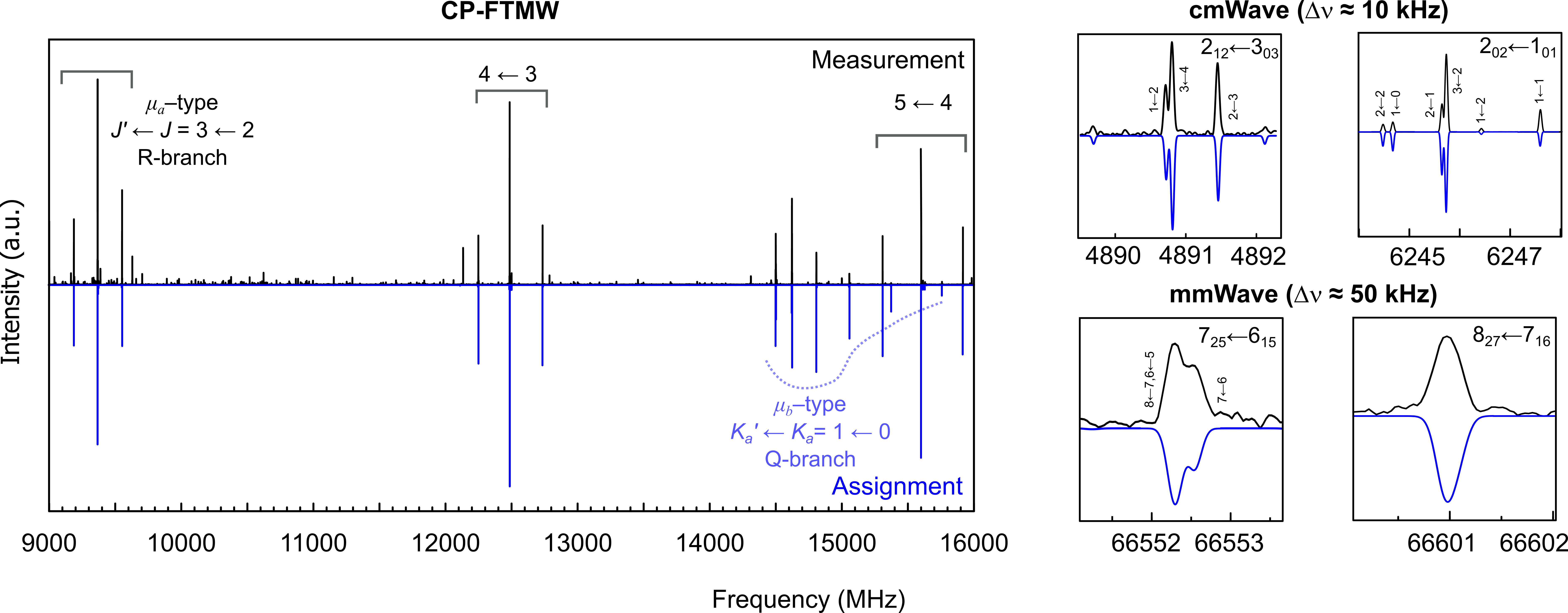}
      \caption{Representative rotational spectrum of 4-oxobutanenitrile assigned to the TC conformer. The main panel shows the broadband CP-FTMW spectrum in the 9-16 GHz range (black), featuring $\mu{\rm _a}$-type R-branch transitions and hyperfine-resolved $\mu{\rm _b}$-type Q-branch features, together with the fitted spectrum (blue). The insets on the right display typical transitions measured with each spectrometer: two in the cm-wave range (top) recorded using the CP-FTMW setup, where the $^{14}$N quadrupole hyperfine structure is clearly resolved, and two in the mm-wave region (bottom) from the FJ-AMMW instrument.}
         \label{fig:spectrum}
\end{figure*}

In recent years, our understanding of the chemical complexity in the interstellar medium (ISM) has advanced dramatically. 
The enhanced sensitivity of modern radiotelescopes -including single-dish facilities such as Yebes 40m, IRAM 30m, and GBT 100m, as well as interferometers like ALMA - has enabled the detection of an unprecedented number of molecules.
Between 2021 and 2025 alone, $>$105 new species were reported in the ISM, accounting for almost one third of
all known interstellar molecules since the first detection in the late 1930s (\citealt{swings1937}).  
These discoveries have been especially prolific in two chemically rich regions: 
the dark molecular cloud TMC-1 and the Galactic Center molecular cloud G+0.693-0.027 (hereafter G+0.693).
In TMC-1, the newly identified molecules include hydrocarbons (some of them aromatic) such as
ortho-benzyne (\ce{o-C6H4}; \citealt{cernicharo2021_benzene}),  
or indene (\ce{c-C9H8}; \citealt{cernicharo2021_indene,burkhardt2021}); and also CN derivatives such as benzonitrile (\ce{c-C6H5CN}; \citealt{mcguire2018}), cyanoacenaphthylene (c-\ce{C12H7CN}; \citealt{Cernicharo2024}), cyanopyrene (c-\ce{C16H9CN}; \citealt{Wenzel2024a, Wenzel2024b}), and cyanocoronene (c-\ce{C24H11CN}; \citealt{wenzel2025}).
Meanwhile, the  G+0.693 cloud has yielded an abundance of complex organic molecules (COMs) containing oxygen (O), nitrogen (N) and sulfur (S).
These include alcohols (\ce{n-CH3CH2CH2OH}; \citealt{jimenez-serra2022}), diols (\ce{Z-1,2-(CHOH)2}; \citealt{rivilla2022a}), carboxylic acids (\ce{HOCOOH}; \citealt{sanz-novo2023}), amides (\ce{NH2COCH2OH}; \citealt{rivilla2023}), amines (\ce{NH2OH}; \citealt{rivilla2020b}; \ce{NH2CH2CH2OH}; \citealt{rivilla2021a}; \ce{CH2CHNH2}; \citealt{zeng2021}), imines (\ce{Z-HCCCHNH}; \citealt{bizzocchi2020}), isocyanates (C$_2$H$_5$NCHO; \citealt{rodriguez-almeida2021b}), thiols (\ce{t-HCOSH} and \ce{C2H5SH}; \citealt{rodriguez-almeida2021a}), and sulfides (\ce{CH3SCH3}; \citealt{sanz-novo2025}).

Many of the species discovered towards G+0.693 are considered potential  precursors in prebiotic chemistry. Within the framework of the RNA-world hypothesis (\citealt{gilbert1986}), which suggests that ribonucleotides were among the first biopolymers to emerge, such compounds play essential roles. 
Laboratory studies have shown that ribonucleotide building blocks can form under plausible prebiotic conditions (\citealt{powner2009,patel2015,becker2019}). 
Their detection in space reinforces the idea that at least part of Earth's early chemical inventory may have originated exogenously. This hypothesis is further supported by the detection of amino acids, nucleobases, and sugars in meteorites (\citealt{cooper2001,elsila2016meteoritic,furukawa2019})), asteroids (\citealt{potiszil2023insights}), and comets (\citealt{altwegg2016}).
This prebiotic material could have been transferred to our young planet via cometary and meteoritic impacts (\citealt{oro1961_comet,cronin1988,chyba1990,chyba1992,cooper2001,pasek2008,glavin2009,pearce2017,marty2017,obrien2018,rubin2019,rivilla2020a}). 
Despite this progress, the exact mechanisms through which complex organics were inherited from the parent molecular cloud remain uncertain. To bridge this gap, it is crucial to identify increasingly complex molecular intermediates in the ISM.  
However, this quest remains challenging, mainly due to the lack of laboratory spectroscopic data.
A strong collaboration between laboratory spectroscopy and observational astronomy is therefore essential. Notably, several detections in G+0.693 arrived closely after their laboratory spectra became available: Z–propynimine (\ce {Z-HCCCHNH}; \citealt{bizzocchi2020}), Z–1,2–ethenediol (Z-1,2-(CHOH)$_2$), isobutene (\ce{(CH3)2C=CH2}; \citealt{fatima2023}) and very recently the tentative identification of 3-hydroxypropanal (\ce{HOCH2CH2CHO}; \citealt{fried2025}).

In this work, we present the rotational spectroscopic study 4-oxobutanenitrile (\ce{HCOCH2CH2CN}; also know as 3-cyanopropionaldehyde), a molecule proposed in  prebiotic chemistry (see Fig. 4d in \citealt{kitadai2018}; and also \citealt{grefenstette2017thesis}) as a precursor of glutamic acid, a proteinogenic amino acid fundamental to modern biochemistry.
The previous identification towards G+0.693 of other nitriles with 4 carbon atoms (the isomeric family \ce{C4H3N}; \citealt{rivilla2022c}), and the discovery of NO-bearing COMs (e.g., \ce{NH2CH2CH2OH}, \ce{CH2OH(O)NH2}; \citealt{rivilla2021,rivilla2023}, respectively), makes this source an excellent candidate to search for \ce{HCOCH2CH2CN}.

After a conformational exploration of the molecule, we characterized its most stable form in the gas phase using broadband laboratory techniques. The derived spectroscopic parameters enabled a targeted astronomical search toward the G+0.693 molecular cloud, a source renowned for its chemical richness (\citealt{zeng2018,rivilla2022c,jimenez-serra2025}). This study exemplifies the synergy between laboratory spectroscopy and radioastronomical observations in advancing our understanding of chemical complexity in space and guiding future detections of molecular precursors relevant to the RNA-world scenario.

\section{Methods}
\label{sec:spectroscopy}

\subsection{Exploration of the Conformational Space: Dihedral Scans and Energetics}
\label{sec:conformational-space}

The conformational landscape of 4-oxobutanenitrile was explored by quantum-chemical methods to support the spectroscopic analysis (\citealt{uriarte2018,calabrese2020}). The molecule exhibits flexibility around two key dihedral angles, C1-C2-C3-C4 ($\theta$) and C2-C3-C4=O ($\phi$), which determine the orientation of its alkyl chain and carbonyl group, respectively. A relaxed two-dimensional potential energy surface (PES) scan was carried out at the B3LYP-D3BJ/def2-TZVPP level using Gaussian 16 (\citealt{frisch2016}) and ORCA (\citealt{neese2020}). This search identified nine energy minima, arising from combinations of gauche, trans, anticlinal, and cis orientations of the two dihedrals (Figure \ref{fig:conformeros}). All conformers were confirmed by harmonic frequency analysis and corrected for zero-point vibrational energy (ZPE).
For the global minimum, TC, the anharmonic approach was also applied by means of the second order vibrational perturbation theory (VPT2) as implemented in Gaussian 16 (\citealt{barone2005, Bloino2012}).

The global minimum corresponds to an extended alkyl backbone ($\theta$ $\approx$ 180$^{\circ}$) and a planar carbonyl group ($\phi$ $\approx$ 0$^{\circ}$). The remaining eight conformers form four pairs of enantiomers. The first pair, G$^{-}$C and G$^{+}$C ($\theta$  $\approx$ $\pm$60$^{\circ}$, $\phi$ $\approx$ 0$^{\circ}$), lies 4.4 kJ mol$^{-1}$ above TC. The other pairs, G$^{-}$A$^{+}$/G$^{+}$A$^{-}$, G$^{-}$A$^{-}$/G$^{+}$A$^{+}$, and TA$^{+}$/TA$^{-}$, span relative energies between approximately 10 and 11 kJ mol$^{-1}$ above the global minimum. Transition states connecting these conformers were located using the Nudged Elastic Band - TS method (\citealt{henkelman2000}). 
The barrier between G$^{-}$C (or the equivalent conformer G$^{+}$C) to TC is approximately 4.1 kJ mol$^{-1}$, while all other interconversion barriers were found to be below 3 kJ mol$^{-1}$.
This shallow landscape enables efficient conformational relaxation under supersonic expansion.

\subsection{Rotational Spectroscopy Using CP-FTMW and FJ-AMMW Techniques}

The rotational spectrum of 4-oxobutanenitrile was recorded using two complementary techniques: chirped-pulse Fourier transform microwave (CP-FTMW) spectroscopy at the University of the Basque Country (\citealt{uriarte2016,uriarte2017}), and free-jet absorption millimeter-wave (FJ-AMMW) spectroscopy at the University of Bologna (\citealt{calabrese2013keto,Calabrese2015} and \citealt{vigorito2018millimeter}).

In the CP-FTMW experiments, the commercial sample was heated to 45$^{\circ}$C and seeded in neon at backing pressures of 5$-$7 bar. The gas mixture was expanded into a high-vacuum chamber ($\sim$10$^{-6}$ mbar) through a pulsed nozzle, generating a supersonic molecular beam. Broadband spectra were recorded in three overlapping frequency windows (2$-$6.5, 7$-$14, and 12$-$18 GHz), each acquired by averaging up to 100,000 free induction decays. The resulting spectra were Fourier transformed to yield high-resolution frequency-domain data.

Measurements in the millimeter-wave region were performed using the FJ-AMMW spectrometer. The sample, maintained at ambient temperature ($\sim$300 K), was entrained in a stream of argon at a stagnation pressure of 19 kPa and expanded through a 0.3 mm pinhole nozzle. The estimated background pressure was $\sim$0.5 Pa, leading to rotational cooling to 5-10 K. Under these conditions, Doppler-limited absorption signals were recorded in the 59.6$-$80.0 GHz range with a typical resolution of 50 kHz. Transitions separated by more than 300 kHz were fully resolved. 
A detailed analysis of selected spectral regions and the assigned transitions is presented in Section \ref{sec:spectral-assignment}.

Only transitions from the global minimim conformer (TC) were observed, and notably no transitions attributable to the next most stable conformers, G$^{+}$C or G$^{-}$C, were detected in either spectrometer. Given the relatively low computed interconversion barrier ($\sim$4.1 kJ mol$^{-1}$ from G$^{+}$C/G$^{-}$C to TC), efficient relaxation toward the global minimum is expected during supersonic expansion. This is consistent with experimental trends reported for other flexible molecules, where barriers below $\sim$4.8 kJ mol$^{-1}$ often lead to full conformational relaxation in the jet (\citealt{ruoff1990}). To investigate the possible stabilization of metastable species, a second CP-FTMW experiment was carried out using helium (5 bar) instead of neon as carrier gas, averaging over 1.5 million free induction decays in the 8$-$16 GHz range. However, the resulting spectrum was virtually identical to that obtained with neon, and again only transitions from the TC conformer were observed. These findings confirm that under the jet-cooled conditions employed, 4-oxobutanenitrile undergoes complete relaxation to its global minimum. The combined use of CP-FTMW and FJ-AMMW spectroscopy, providing broad spectral coverage and high-resolution hyperfine structure, enabled the full and unambiguous characterization of the TC conformer.

\begin{table}
\caption[c]{Spectroscopic datasets and their contributions to the final fit.}
\label{table:summary}
\centering
\begin{center}
\begin{footnotesize}
\setlength{\tabcolsep}{5pt}
\begin{tabular}{c r r r r r}
\hline\hline
Experiment & Lines$^{(a)}$   & $\sigma$$^{(b)}$ & $J_{\rm up}$  & $K_{\rm a}$ & Frequency \\
 &   & (kHz)  &  range &  range & range (GHz)\\
\cline{1-6}
CP-FTMW   &  87                        & 12  &  1$-$5  & 0$-$2 & 3.122$-$17.502\\  
FJ-AMMW   &  21 & 34  &  6$-$20 & 0$-$2 & 59.974$-$77.666\\    
\hline
\end{tabular}
\end{footnotesize}
\end{center}
{${(\rm a)}$: considering hyperfine structure;
${(\rm b)}$:} $\sigma$ is the root mean square deviation of the fit.
\end{table}

\begin{table}
\caption[c]{Calculated and experimental spectroscopic parameters of the observed conformer TC of 4-oxobutanenitrile, derived from a global fit of its rotational transitions recorded in the 2$-$18 GHz and 59.6$-$80 GHz ranges.}
%\centering
\label{table:spectroscopic}
\begin{center}
\begin{footnotesize}
\setlength{\tabcolsep}{10pt}
\begin{tabular}{l r r}
\hline\hline
Parameter &  Theory & Experiment \\
\hline
$A_{\rm 0}$  (MHz)                               &  16103.08  &   16000.7925(30)       \\
$B_{\rm 0}$ (MHz)                                &  1608.79  &   1622.40376(40)     \\
$C_{\rm 0}$  (MHz)                               &  1490.00  &   1500.80587(40)     \\
$\frac{3}{2}\chi_{\rm aa}$ (MHz)                 &  -6.60    &   -5.8503(91)        \\
$\frac{1}{4}(\chi_{\rm bb}-\chi_{\rm cc})$ (MHz) &  -0.06    &   -0.0471(35)        \\
$D_{\rm J}$       (kHz)                          &   0.193   &    0.19471(64)       \\
$D_{\rm JK}$   (kHz)                             &  -1.144   &   -0.925(76)         \\
$D_{\rm K}$   (kHz)                              &  36.955   &   36.8(15)           \\
$d_{1}$   (kHz)                                  &  -0.021   &   -0.02012(64)       \\
$d_{2}$   (kHz)                                  &  -0.001   &   -0.00254(75)       \\
$|\mu_{\rm a}|$ (D)                              &    2.60   &     Y                \\
$|\mu_{\rm b}|$ (D)                              &    1.54   &     Y                \\
$|\mu_{\rm c}|$ (D)                              &      0    &     N                \\
\hline
\end{tabular}
\end{footnotesize}
\end{center}
NOTE -
$A_{\rm 0}$, $B_{\rm 0}$, and $C_{\rm 0}$ are the rotational constants in the vibrational ground state (the corresponding equilibrium rotational constants are: $A_{\rm e}$=16250.00, $B_{\rm e}$=1617.95, $C_{\rm e}$=1498.13 MHz);
$D_{\rm J}$, $D_{\rm JK}$, $D_{\rm K}$, $d_{\rm 1}$ and $d_{\rm 2}$ are the quartic centrifugal distortion constants;
$\chi_{\rm aa}$, $\chi_{\rm bb}$ and $\chi_{\rm cc}$ are the diagonal elements of the $^{14}$N nuclear quadrupole coupling tensor;
$|\mu_{\rm a}|$, $|\mu_{\rm b}|$ and $|\mu_{\rm c}|$ are the absolute values of the electric dipole moment components along the principal inertial axes ($\mu_{\rm c}$ is zero by symmetry), Y or N denotes that such type of transitions have or have not been observed.
\end{table}

\begin{table}
\caption{Rotational and vibrational partition functions for 4-oxobutanenitrile at different temperatures.}
\label{table:partition}
%\centering
\begin{center}
\setlength{\tabcolsep}{10pt}
\begin{tabular}{ r r r }
\hline\hline
$T$ (K) &  $Q_{\text{rot}}$ & $Q_{\text{vib}}$ \\
\hline
   300.000    & 421710.52 & 27.6 \\ 
   225.000    & 273819.18 & 10.7 \\
   150.000    & 148996.77 &  4.0 \\
    75.000    & 52665.66  &  1.6 \\
    37.500    & 18622.36  &  1.1 \\
    18.750    & 6587.55   &  1.0 \\
     9.375    & 2331.92   &  1.0 \\
     7.400    & 1636.43   &  1.0 \\
     3.000    & 424.42    &  1.0 \\     
\hline
\end{tabular}
\end{center}
\end{table}

\subsection{Spectral Assignment and Conformer Identification}
\label{sec:spectral-assignment}

The rotational spectra of 4-oxobutanenitrile revealed the presence of a single conformer, whose assignment was based on the observation of characteristic $\mu_{\rm a}-$type R-branch transitions such as 3$_{0,3}$ $\leftarrow$ 2$_{0,2}$, 4$_{0,4}$ $\leftarrow$ 3$_{0,3}$, and 5$_{0,5}$ $\leftarrow$ 4$_{0,4}$. All transitions exhibited hyperfine structure arising from the $^{14}$N nuclear quadrupole coupling ($I$ = 1). A total of 87 hyperfine components were assigned in the 2$-$18 GHz region, including  $\mu_{\rm b}-$type transitions and lines up to $J_{\rm up}$ = 5 and $K_{\rm a}$ = 2 (see Table \ref{table:summary}).
In the 59.974$-$77.666 GHz domain, 16 additional rotational transitions (21 considering hyperfine structure) were identified using FJ-AMMW spectroscopy (Table \ref{table:summary}). These included higher-$J$ $\mu_{\rm a}-$ and $\mu_{\rm b}-$type transitions, extending up to $J_{\rm up}$ = 20 and $K_{\rm a}$ = 2. 
Portion of the spectra in significant regions are shown in Figure \ref{fig:spectrum}. In total, 40 rotational transitions were measured (108 considering hyperfine structure), including 22 $\mu_{\rm a}$-type R-branch transitions up to $J_{\rm up}$ and $K_{\rm a}$ = 2, 2 $\mu_{\rm b}$-type P-branch up to $J_{\rm up}$ = 3 and 4 $\mu_{\rm b}$-type Q-branch transitions up to $J$ = 4,
%with low $J_{\rm up}$ values, 
and 12 $\mu_{\rm b}$-type R-branch transitions up to $J_{\rm up}$ = 12 and $K_{\rm a}$=2. The determined spectroscopic constants allow the accurate prediction of $\mu_{\rm a}$-type R-branch transitions with $K_{\rm a}$ = 0, 1, 2 up to 50 GHz and $E_{\rm up}$ up to 23.1 K. The frequency uncertainties, in velocity units, are in the range 0.05$-$0.1 km s$^{-1}$. This enables the search towards the Galactic Center G+0.693-0.027 molecular cloud (see Section \ref{sec:observations}), which exhibits much larger linewidths
($\sim$15-20 km s$^{-1}$; \citealt{requena-torres_largest_2008,rivilla2022c}), and where the rotational population is concentrated in the lowest energy levels due to low $T_{\rm ex}$. The transitions measured here can also assist the identification of Galactic cold clouds with typical linewidths of $\sim$0.5 km s$^{-1}$ and $T_{\rm ex}\sim$10 K (e.g., \citealt{agundez2023}).

For increasing values of $J$ and $K$, and higher frequencies (up to $J$=99 and 300 GHz), the frequency predictions naturally show larger uncertainties. For several transitions, including those reaching upper state energies of $E_{\rm up}\sim$1130 K, the uncertainties still remain $<$ 1 km s$^{-1}$. This accuracy is sufficient to enable searches also towards hot cores, where typical linewidths are 3$-$8 km$^{-1}$ (e.g., \citealt{cesaroni1998,Rivilla:2017ce,colzi2021,belloche2025}). However, for transitions whose predicted uncertainties exceed 1 km s$^{-1}$, improved frequency values would undoubtedly benefit from additional measurements in the (sub)millimetre wavelength range.

We provide the full line list, including frequency uncertainties, as supplementary material so that the applicability of the predicted frequencies to different kinds of astronomical sources can be easily evaluated. 
Nevertheless, these uncertainties must be regarded as lower limit estimates, because they only account for the errors of the fitted spectroscopic parameters. They do not include additional contributions such as higher order centrifugal distortion terms, which would be required to obtain more reliable predictions at higher frequencies.

The observed spectrum corresponds unambiguously to the TC conformer, the global minimum identified in our quantum chemical calculations. 
The theoretical and experimentally derived rotational, quartic centrifugal distortion, and nuclear quadrupole constants are presented in Table \ref{table:spectroscopic}.
All observed lines (Table \ref{table:rot_lines}) were fitted using Watson's S-reduced Hamiltonian in the $I^{\rm r}$ representation (\citealt{watson1977}).
Its assignment is further supported by the detection of $^{13}$C, $^{15}$N, and $^{18}$O isotopologues in natural abundance ($\sim$1.1$\%$, $\sim$0.4$\%$, and $\sim$0.2$\%$, respectively), with measured transitions (Tables \ref{table:rot_lines13C1}-\ref{table:rot_lines15N}) matching those predicted for TC.
The rotational constants and nuclear quadrupole coupling constants for the isotopologues are given in the Appendix (Table \ref{table:spectroscopic_iso}).
Moreover, in agreement with the $C_s$ symmetry of the TC conformer, no $\mu_{\rm c}$-type transition lines were observed.

The rotational partition function ($Q_{\rm rot}$) at different temperatures, reported in Table \ref{table:partition}, was evaluated by summing the Boltzmann factors over the energy levels in the ground vibrational state. 
We used the SPCAT program (\citealt{Pickett1991}) to perform this numerical summation, employing the spectroscopic constants from Table \ref{table:spectroscopic} setting maximum frequency 1 THz and $J$ value 200.
The vibrational partition function ($Q_{\rm vib}$), also shown in Table \ref{table:partition}, was calculated using the vibrational mode frequencies obtained through the anharmonic approximation (VPT2), employing the B3LYP-D3BJ method and the def2-TZVP basis set, as described in Section \ref{sec:conformational-space}.

The spectroscopic data for all isotopologues, including fitting and prediction files, is offered as supplementary material, and additionally it have been posted into the Strasbourg astronomical Data Center (CDS).

\section{Interstellar search towards the G+0.693$-$0.027 molecular cloud}
\label{sec:observations}

\begin{figure*}
     \centering
    \includegraphics[width=0.95\linewidth]{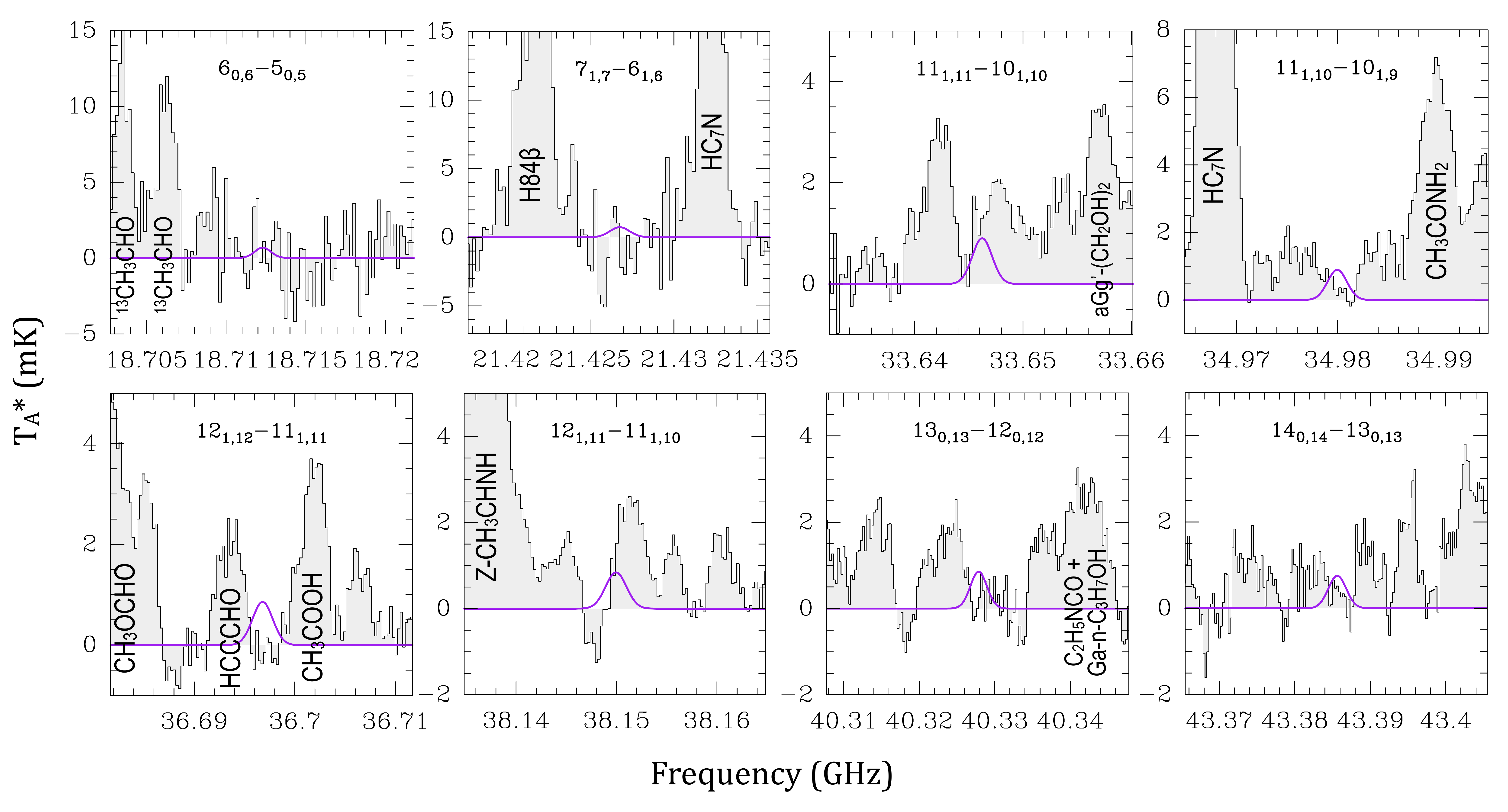}
      \caption{Transitions of \ce{HCOCH2CH2CN} used to compute the column density upper limit towards G+0.693 (see text). Each panel shows the contribution of three transitions that are not spectrally resolved (see Table \ref{table:observations}). The black line and gray histogram show the observed spectrum, the purple curve corresponds to the LTE synthetic model of \ce{HCOCH2CH2CN} using the  column density upper limit derived in this work. Transitions of other species identified in the survey are labeled.  
      %and the blue curve indicate the emission of all the molecules previously identified towards the source.
      }
         \label{fig:g0693}
\end{figure*}

To evaluate the presence of 4-oxobutanenitrile in the ISM, we conducted a targeted search towards the G+0.693 molecular cloud, located in the Sgr B2 region of the Galactic Center. This source is well known for its chemical richness and has been a prolific hunting ground for new molecular detections in recent years (\citealt{rivilla2019b,rivilla2020b,rivilla2021a,rivilla2021b,rivilla2022a,rivilla2022b,rivilla2023,rodriguez-almeida2021a,rodriguez-almeida2021b,zeng2021,jimenez-serra2022,sanz-novo2023,sanz-novo2024a}).
Our analysis was based on high-sensitivity, broadband spectral surveys obtained with the NRAO 100m Green Bank telescope (GBT) in West Virginia (USA), and the Yebes 40m telescope (Guadalajara, Spain).
These observations cover the spectral regions where the strongest transitions of \ce{HCOCH2CH2CN} at the typical excitation temperatures of G+0.693 ($\sim$5$-$20 K) are predicted. 
Further observational details, including frequency coverage, spectral resolution and beam sizes, are given in \citet{zeng2018} and \citet{rivilla2023}.
Given the extended nature of the molecular emission in G+0.693 (e.g., \citealt{brunken_interstellar_2010, Jones2012,Li2020,Santa-Maria2021, Colzi2024}), all intensities were expressed in the $T_{\mathrm{A}}^{\ast}$ scale.

\begin{table}
\centering
\tabcolsep 3pt
\caption{Selected transitions of \ch{HCOCH2CH2CN} used to compute the column density upper limit in the G+0.693$-$0.027 molecular cloud (see Figure \ref{fig:g0693}). Each group of three entries corresponds to unresolved hyperfine components of a rotational transition. Frequencies, intensities (at 300 K), and upper-state energies were calculated using the  spectroscopic parameters derived in this work. Numbers in parenthesis in the frequency values give the uncertainty on the last digits.}
\label{table:observations}
\begin{tabular}{cccccccccc}
\hline\hline
 Frequency & Transition & log \textit{I}& \textit{g}$\mathrm{_u}$ & $E$$\mathrm{_{up}}$  \\ 
 (GHz) & $J_{K_{\rm a},K_{\rm c}}$ $F$  &  (nm$^2$ MHz) &  &  (K)   \\
\hline
18.7122133(34) & 6$_{0,6}$ 5 -- 5$_{0,5}$ 4         & --6.740   & 11 & 3.1  \\ 
18.7122421(34) & 6$_{0,6}$ 6 -- 5$_{0,5}$ 5         & --6.666   & 13 & 3.1 \\ 
18.7122578(34) & 6$_{0,6}$ 7 -- 5$_{0,5}$ 6         & --6.591   & 15 & 3.1 \\ 
\hline      
21.4267372(48) & 7$_{1,7}$ 7 -- 6$_{1,6}$ 6         & --6.489   & 15 & 4.8  \\ 
21.4267461(48) & 7$_{1,7}$ 6 -- 6$_{1,6}$ 5         & --6.552   & 13 & 4.8  \\ 
21.4267730(48) & 7$_{1,7}$ 8 -- 6$_{1,6}$ 7         & --6.426   & 17 & 4.8  \\ 
\hline 
33.6463327(89) & 11$_{1,11}$ 10  -- 10$_{1,10}$ 9   & --5.937   & 21 & 10.4  \\
33.6463327(89) & 11$_{1,11}$ 11  -- 10$_{1,10}$ 10  & --5.898   & 23 & 10.4 \\
33.6463445(89) & 11$_{1,11}$ 12  -- 10$_{1,10}$ 11  & --5.858   & 25 & 10.4 \\
\hline   
34.9798050(83) & 11$_{1,10}$ 10 -- 10$_{1,9}$  9    & --5.864   & 21 & 10.8 \\ 
34.9798054(83) & 11$_{1,10}$ 11 -- 10$_{1,9}$ 10    & --5.904   & 23 & 10.8 \\ 
34.9798166(83) & 11$_{1,10}$ 12 -- 10$_{1,9}$ 11    & --5.824   & 25 & 10.8 \\ 
\hline
36.6969490(104) & 12$_{1,12}$ 11  -- 11$_{1,11}$ 10  & --5.822   & 23 & 12.1  \\
36.6969494(104) & 12$_{1,12}$ 12  -- 11$_{1,11}$ 11  & --5.786   & 25 & 12.1 \\
36.6969590(104) & 12$_{1,12}$ 13  -- 11$_{1,11}$ 12  & --5.749   & 27 & 12.1 \\
\hline  
38.1498001(96) & 12$_{1,11}$ 11 -- 11$_{1,10}$ 10   & --5.789   & 23 & 12.6 \\ 
38.1498001(96) & 12$_{1,11}$ 12 -- 11$_{1,10}$ 11   & --5.753   & 25 & 12.6\\ 
38.1498095(96) & 12$_{1,11}$ 13 -- 11$_{1,10}$ 12   & --5.716   & 27 & 12.6\\ 
\hline 
40.3279643(100) & 13$_{0,13}$ 12  -- 12$_{0,12}$ 11  & --5.701   & 25 & 13.6  \\
40.3279678(100) & 13$_{0,13}$ 13  -- 12$_{0,12}$ 12  & --5.668   & 27 & 13.6 \\ 
40.3279731(100) & 13$_{0,13}$ 14  -- 12$_{0,12}$ 13  & --5.634   & 29 & 13.6 \\ 
\hline 
43.3857819(118) & 14$_{0,14}$ 13 -- 13$_{0,13}$ 12   & --5.605   & 27 & 15.7 \\
43.3857846(118) & 14$_{0,14}$ 14 -- 13$_{0,13}$ 13   & --5.575   & 29 & 15.7 \\ 
43.3857895(118) & 14$_{0,14}$ 15 -- 13$_{0,13}$ 14   & --5.543   & 31 & 15.7 \\ 
\hline
\end{tabular}
\vspace*{-2.5ex}
\end{table}

We implemented the newly derived spectroscopic parameters of \ce{HCOCH2CH2CN} into the MADCUBA package{\footnote{Madrid Data Cube Analysis on ImageJ is a software developed at the Center of Astrobiology (CAB) in Madrid; http://cab.inta-csic.es/madcuba/}} (\citealt{martin2019}). We generated the Local Thermodynamic Equilibrium (LTE) synthetic spectra using the SLIM (Spectral Line Identification and Modeling) tool of MADCUBA, which were compared to the observational data. We note that, given the low excitation temperatures of the molecules in G+0.693, we did not include the vibrational contribution to the partition function ($Q_{\rm vib}$) in this case.
No signals attributable to \ce{HCOCH2CH2CN} were found.
To derive an upper limit to the column density, we selected the most intense and least blended transitions predicted by the LTE model.
These transitions (all $\mu_{\rm a}$-type R-branch) fall within the frequency ranges highlighted in Figure \ref{fig:g0693}, with root mean square (rms) noise levels of $\sim$2.3 mK for the GBT data, and 0.4$-$1 mK for the Yebes data.
The selected transitions appear as groups of three unresolved hyperfine components (see Table \ref{table:observations}). 
The LTE simulations assumed excitation parameters consistent with previous fits to other nitrile- and formyl-bearing species in the same source, specifically $T_{\rm ex}$=7.4 K,  v$_{\rm LSR}$=68 km s$^{-1}$ and FWHM=19.2 km s$^{-1}$ (\citealt{rivilla2022c}).
Using this model, we computed the maximum column density ($N$) consistent with the absence of detectable signal, yielding a conservative upper limit of $N<$ 4 $\times$ 10$^{12}$ cm$^{-2}$. This corresponds to a molecular abundance of $<$ 2.9 $\times$ 10$^{-11}$ relative to H$_2$, adopting $N$(H$_{2}$)=1.35$\times$10$^{23}$ cm$^{-2}$ (\citealt{martin_tracing_2008}). 
We note that this abundance upper limit is significantly lower, by a factor of 9, than that derived previously for the simpler 3-oxopropanenitrile
(\ce{HCOCH2CN}; \citealt{rivilla2022c}).

\section{Discussion: astronomical implications}

\begin{figure}
     \centering
    \includegraphics[width=1.0\linewidth]{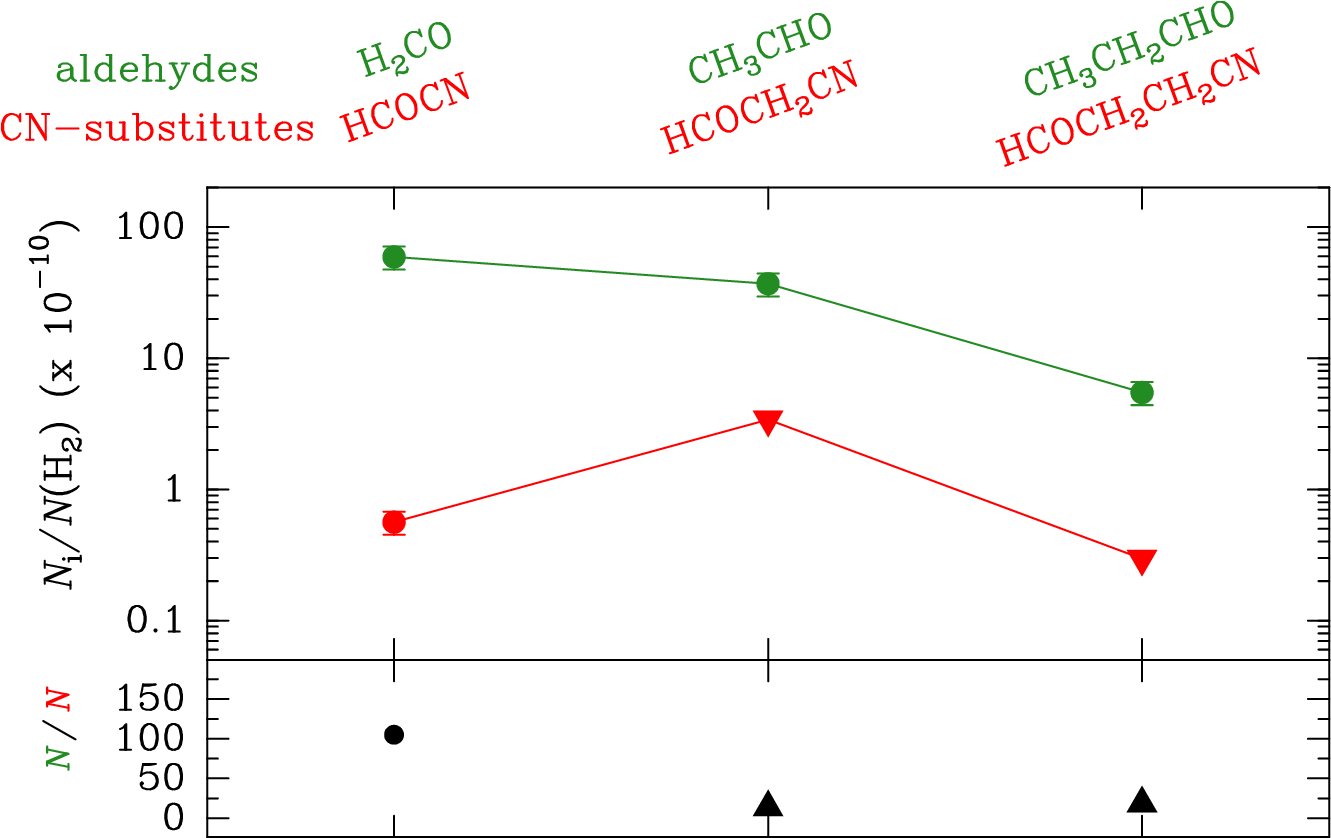}
      \caption{Upper panel: molecular abundances of aldehydes detected towards G+0.693 $-$ formaldehyde, acetaldehyde and propionaldehyde (propanal) $-$ from \citet{rivilla2020b,jimenez-serra2020,sanz-novo2022}; compared to those of their CN$-$substitutes $-$ oxoacetonitrile, 3-oxopropanenitrile and 4-oxobutanenitrile $-$ from \citet{rivilla2022c} and this work. An uncertainty of 20$\%$ for the molecular abundances has been assumed. Upper limits are denoted with downward-pointing triangles. Lower panel: Column density molecular ratio of the aldehyde/CN$-$substitute pairs. Lower limits are denoted with upward-pointing triangles.}
         \label{fig:ratios-adehyde}
\end{figure}

The derived upper limit for the molecular abundance of   \ce{HCOCH2CH2CN} (11 atoms) indicates that it is significantly less abundant than the largest species detected so far towards G+0.693, which are ethanolamine (\ce{NH2CH2CH2OH}, 11 atoms; \citealt{rivilla2021a}), and n$-$propanol (\ce{CH3CH2CH2OH}, 12 atoms; \citealt{jimenez-serra2022}), whose abundances are of several 10$^{-10}$. 
To place the derived upper limit in context, we compared with other structurally similar species.
\ce{HCOCH2CH2CN} can be seen as the CN$-$substitute of the interstellar molecule propionaldehyde (\ce{CH3CH2CHO}, also known as propanal). Similarly, the CN$-$substitutes of two abundant interstellar aldehydes, formaldehyde (\ce{H2CO}) and acetaldehyde (\ce{CH3CHO}), are oxoacetonitrile (\ce{HCOCN}) and 3-oxopropanenitrile (\ce{HCOCH2CN}). 
To test observationally how efficient is the $-$CN substitution of aldehydes in the ISM, we compare in Figure \ref{fig:ratios-adehyde} the abundances derived towards G+0.693 of these aldehydes (from \citealt{rivilla2020b,jimenez-serra2020,sanz-novo2022}) with the values (or upper limits) of their CN$-$substitutes (from \citealt{rivilla2022c} and this work). 
The abundances of aldehydes decrease with increasing complexity, as observed in other families of molecules towards G+0.693 (alcohols, isocyanates, thiols; \citealt{jimenez-serra2022,rodriguez-almeida2021b,rodriguez-almeida2021a}). Assuming that the \ce{H2CO}/\ce{HCOCN} ratio of $\sim$100 (Figure \ref{fig:ratios-adehyde}) is similar for the more complex aldehyde/CN$-$substitute pairs, the expected abundances for \ce{HCOCH2CN} and \ce{HCOCH2CH2CN} would be at least one order of magnitude lower than the upper limits derived (see Figure \ref{fig:ratios-adehyde}). These low abundances are well below the current sensitivity limits of the astronomical observations.

To infer the expected abundance of \ce{HCOCH2CH2CN} %(and \ce{HCOCH2CN}) 
in the ISM, and therefore to open the possibility for its future detection, we need to consider the viability of possible chemical routes working at interstellar conditions. 
There are not, to our knowledge, theoretical and/or experimental works devoted to the interstellar synthesis of \ce{HCOCH2CH2CN}.
We discuss here some possible chemical pathways, based on previous studies of simpler CN$-$substitutes of aldehydes. 
The quantum chemical calculations by \citet{tonolo2020} showed that \ce{HCOCN} can be efficiently formed through the gas-phase reaction
between \ce{H2CO} and the CN radical.
Therefore, one might think that the reactions of CN with \ce{CH3CHO} and \ce{CH3CH2CHO} can form \ce{HCOCH2CN} and \ce{HCOCH2CH2CN}, respectively. 
However, \citet{tonolo2020} also showed that CN + \ce{CH3CHO} has two possible product paths: i) \ce{HCOCN} + \ce{CH3}, and ii) H + acetyl cyanide (\ce{CH3C(O)CN}), an isomer of \ce{HCOCH2CN}. Acetyl cyanide is formed through the CN attack onto the unsaturated carbon of \ce{CH3CHO}, while the formation of \ce{HCOCH2CN} would need the CN attack onto the saturated carbon of the methyl group, which is not favored according to  \citet{tonolo2020} calculations. It was also demonstrated in other systems that the CN attack onto unsaturated carbons are dominant (e.g., \citealt{balucani2000}). It is therefore unlikely that the gas-phase reaction between CN and \ce{CH3CH2CHO} conducts to the formation of \ce{HCOCH2CH2CN}.

A possible alternative is the reaction of CN with propylene oxide, \ce{CH3(O)CHCH3}. The analogous reaction with ethylene oxide (\ce{c-C2H4O}) produces \ce{HCOCH2CN} + H (\citealt{alessandrini2021}), although the main reaction pathway is 2-oxiranyl radical + HCN. Since propylene oxide has been identified in the ISM (\citealt{mcguire2016}), we encourage new theoretical calculations to study in detail whether its reaction with CN can form \ce{HCOCH2CH2CN} at a significant rate.

\begin{figure*}
     \centering
    \includegraphics[width=0.45\linewidth]{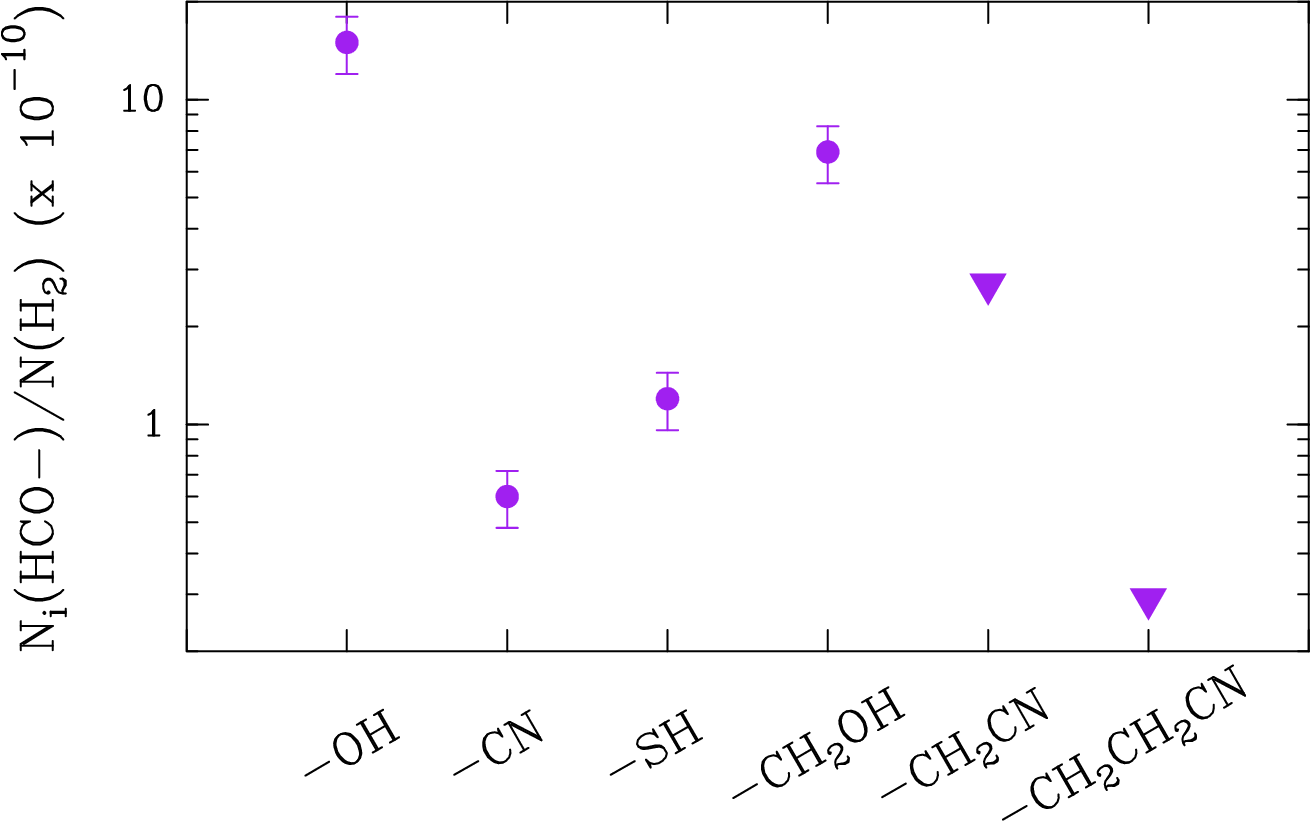}
\hspace{1cm}
%    \vskip0.3cm
    \includegraphics[width=0.45\linewidth]{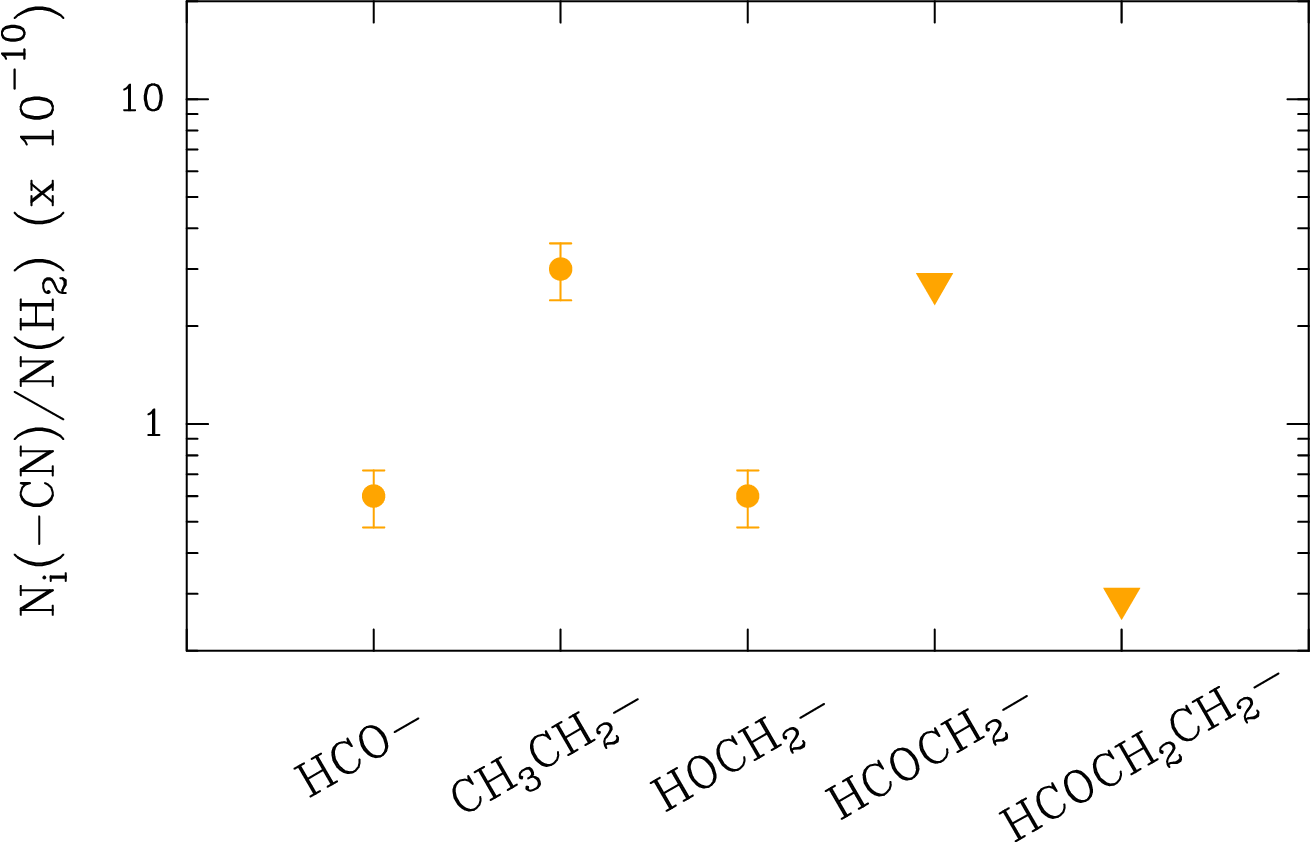}
      \caption{Molecular abundances derived towards the G+0.693 molecular cloud of molecules containing the formyl group ($-$HCO, left panel), and the nitrile group ($-$CN, right panel), based on the values reported in this work and \citet{zeng2018,rodriguez-almeida2021a,rivilla2022a,rivilla2022c}. An uncertainty of 20$\%$ for the molecular abundances has been assumed. Upper limits are denoted with downward-pointing triangles.}
         \label{fig:abundances}
\end{figure*}

Within a broader view of the chemical content in G+0.693, we also compared the upper limits derived for \ce{HCOCH2CH2CN} with other nitrile- and formyl-containing species detected (or searched for) towards G+0.693 in previous works. 
These include formic acid (\ce{HCOOH}, $trans$-conformer; \citealt{rodriguez-almeida2021a}), cyanoformaldehyde (\ce{HCOCN}; \citealt{rivilla2022c}), thioformic-acid (\ce{HCOSH}, $trans$-conformer; \citealt{rodriguez-almeida2021a}), glycolaldehyde (\ce{HCOCH2OH}; \citealt{rivilla2022a}), ethylcyanide (\ce{CH3CH2CN}; \citealt{zeng2018}), glycolonitrile (\ce{HOCH2CN}; \citealt{rivilla2022c}), and 3-oxopropanenitrile
(\ce{HCOCH2CN)} (\citealt{rivilla2022c}). 
As shown in Figure \ref{fig:abundances}, the abundance constraint for \ce{HCOCH2CH2CN} is the lowest. This underscores again the importance of even more sensitive observations to detect the next tier of interstellar complexity, especially molecules with plausible links to prebiotic chemistry. Completing the inventory of -CN and HCO-bearing compounds in chemically rich regions like G+0.693 will help refine models for the emergence of molecular complexity in space.

\section{Conclusions and outlook}
\label{sec:conclusions}

We have investigated the rotational spectra of  4-oxobutanenitrile (\ch{HCOCH2CH2CN}), a proposed precursor of glutamic acid, an amino acid essential for protein biosynthesis. 
Through quantum chemical calculations, we mapped its conformational space, identifying nine stable minima. The global minimum conformer was confirmed experimentally, and its rotational spectrum was recorded in the 2$-$18 GHz and 59.6$-$80 GHz regions with high resolution, enabling its  astronomical search towards the G+0.693-0.027 molecular cloud, located in the Sgr B2 complex.
The molecule was not detected, and we derived a stringent upper limit to its column density of $N<$ 4 $\times$10$^{12}$ cm$^{-2}$, corresponding to a fractional abundance of $<$ 2.9 $\times$10$^{-11}$ relative to H$_2$. This upper limit is significantly lower than the derived abundances of chemically similar species containing $-$HCO and $-$CN groups, and also lower than those of the most complex species detected so far towards this cloud.
We have discussed the viability of several chemical pathways to form this molecule in the ISM, aiming to pave the way for a future detection.
Our findings reinforce the need to enhance the sensitivity of current radioastronomical instrumentation to access a new generation of elusive molecular species. Such efforts are essential to expand our inventory of prebiotic molecules in the interstellar medium and to deepen our understanding of the chemical steps that may lead toward the emergence of life.

\section*{Acknowledgements}

We are very grateful to the Yebes 40 m and IRAM 30 m telescope staff for their precious help during the different observing runs. The 40 m radio telescope at Yebes Observatory is operated by the Spanish Geographic Institute (IGN; Ministerio de Transportes, Movilidad y Agenda Urbana). 
V.M.R. and I.J-S. acknowledge support from the grant No. PID2022-136814NB-I00 by the Spanish Ministry of Science, Innovation and Universities/State Agency of Research MCIU/AEI/ 10.13039/501100011033 and by ERDF, UE.
V.M.R. also acknowledges support from the grant PID2022-136814NB-I00 by the Spanish Ministry of Science, Innovation and Universities/State Agency of Research MICIU/AEI/10.13039/501100011033 and by ERDF, UE;  the grant RYC2020-029387-I funded by MICIU/AEI/10.13039/501100011033 and by "ESF, Investing in your future", and from the Consejo Superior de Investigaciones Cient{\'i}ficas (CSIC) and the Centro de Astrobiolog{\'i}a (CAB) through the project 20225AT015 (Proyectos intramurales especiales del CSIC); and from the grant CNS2023-144464 funded by MICIU/AEI/10.13039/501100011033 and by “European Union NextGenerationEU/PRTR”.
W.S. acknowledges a post-doctoral fellowship under the MUR PRIN national project (grant n. J43C21000050001).
E.J.C. and F.J.B. acknowledge financial support from the Spanish Ministry of Science, Innovation and Universities (MICIU/AEI, project PID2023-147698NB-I00) and from the Basque Government (project IT1491-22).
A.I. acknowledges a postdoctoral fellowship from the Basque Government. The authors also acknowledge the use of computational resources provided by SGIker (UPV/EHU) and CESGA.
S. M. and A.M acknowledge financial support from the Italian Space Agency, ASI, and the Ministry of University and Research, MUR, through the Space It Up project funded under contract n. 2024-5-E.0 - CUP n. I53D24000060005.
The authors also thank the University of Bologna for financial support through the RFO project.
I.J.-S. acknowledges funding from the ERC CoG grant OPENS (project number 101125858) funded by the European Union.

%%%%%%%%%%%%%%%%%%%%%%%%%%%%%%%%%%%%%%%%%%%%%%%%%%
\section*{Data Availability}

The observational data of the G+0.693-0.027 spectral survey used in this article will be shared on reasonable request to vrivilla@cab.inta-csic.es.
The spectroscopic data, including fitting and prediction files, is offered as supplementary material, and additionally have been posted into the Strasbourg astronomical Data Center (CDS).



\bibliographystyle{mnras}
\bibliography{references,library,rivilla}

%\begin{thebibliography}{99}
%\end{thebibliography}

%%%%%%%%%%%%%%%%%%%%%%%%%%%%%%%%%%%%%%%%%%%%%%%%%%

%%%%%%%%%%%%%%%%% APPENDICES %%%%%%%%%%%%%%%%%%%%%
\appendix

\section{Rotational spectroscopy data} 
%Spectroscopic parameters of the $^{13}$C, $^{15}$N, and $^{18}$O isotopologues of 4-oxobutanenitrile

\begin{table*}
\caption[c]{Measured rotational transition frequencies and corresponding deviations from the model of the TC conformer of 4-oxobutanenitrile.}
%\centering
\label{table:rot_lines}
\begin{center}
\begin{footnotesize}
\setlength{\tabcolsep}{4pt}
\begin{tabular}{l l r r c l l r r}
\hline\hline
$J'_{K'_{\rm a},K'_{\rm c}}$ $F'$&$J"_{K"_{\rm a},K"_{\rm c}}$ $F"$& freq. (MHz) & o.-c. (MHz) &~~~~&
$J'_{K'_{\rm a},K'_{\rm c}}$ $F'$&$J"_{K"_{\rm a},K"_{\rm c}}$ $F"$& freq. (MHz) & o.-c. (MHz) \\
\hline
1	$_{	0	,	1	}$	1	&	0	$_{	0	,	0	}$	1	&	3122.233	&	-0.001	&&	1	$_{	1	,	0	}$	1	&	1	$_{	0	,	1	}$	2	&	14500.270	&	0.002	\\
1	$_{	0	,	1	}$	2	&	0	$_{	0	,	0	}$	1	&	3123.406	&	0.002	&&	1	$_{	1	,	0	}$	2	&	1	$_{	0	,	1	}$	1	&	14500.825	&	0.001	\\
1	$_{	0	,	1	}$	0	&	0	$_{	0	,	0	}$	1	&	3125.162	&	0.003	&&	1	$_{	1	,	0	}$	1	&	1	$_{	0	,	1	}$	1	&	14501.442	&	0.005	\\
2	$_{	0	,	2	}$	1	&	1	$_{	0	,	1	}$	0	&	6244.671	&	0.000	&&	2	$_{	1	,	1	}$	2	&	2	$_{	0	,	2	}$	1	&	14620.883	&	0.002	\\
2	$_{	0	,	2	}$	2	&	1	$_{	0	,	1	}$	2	&	6244.475	&	0.001	&&	2	$_{	1	,	1	}$	2	&	2	$_{	0	,	2	}$	3	&	14621.576	&	-0.002	\\
2	$_{	0	,	2	}$	3	&	1	$_{	0	,	1	}$	2	&	6245.731	&	0.003	&&	2	$_{	1	,	1	}$	1	&	2	$_{	0	,	2	}$	1	&	14621.811	&	0.001	\\
2	$_{	0	,	2	}$	2	&	1	$_{	0	,	1	}$	1	&	6245.646	&	0.001	&&	2	$_{	1	,	1	}$	3	&	2	$_{	0	,	2	}$	3	&	14622.171	&	-0.003	\\
2	$_{	0	,	2	}$	1	&	1	$_{	0	,	1	}$	2	&	6246.427	&	0.002	&&	2	$_{	1	,	1	}$	2	&	2	$_{	0	,	2	}$	2	&	14622.831	&	-0.001	\\
2	$_{	0	,	2	}$	1	&	1	$_{	0	,	1	}$	1	&	6247.598	&	0.002	&&	2	$_{	1	,	1	}$	3	&	2	$_{	0	,	2	}$	2	&	14623.429	&	0.001	\\
2	$_{	1	,	2	}$	2	&	1	$_{	1	,	1	}$	1	&	6123.849	&	0.004	&&	2	$_{	1	,	1	}$	1	&	2	$_{	0	,	2	}$	2	&	14623.759	&	-0.002	\\
2	$_{	1	,	2	}$	2	&	1	$_{	1	,	1	}$	2	&	6124.403	&	0.002	&&	3	$_{	1	,	2	}$	3	&	3	$_{	0	,	3	}$	2	&	14805.824	&	0.001	\\
2	$_{	1	,	2	}$	1	&	1	$_{	1	,	1	}$	1	&	6124.873	&	0.006	&&	3	$_{	1	,	2	}$	3	&	3	$_{	0	,	3	}$	4	&	14806.271	&	-0.007	\\
2	$_{	1	,	2	}$	3	&	1	$_{	1	,	1	}$	2	&	6125.064	&	0.005	&&	3	$_{	1	,	2	}$	2	&	3	$_{	0	,	3	}$	2	&	14807.084	&	-0.013	\\
2	$_{	1	,	2	}$	1	&	1	$_{	1	,	1	}$	0	&	6126.264	&	0.005	&&	3	$_{	1	,	2	}$	4	&	3	$_{	0	,	3	}$	4	&	14807.231	&	0.009	\\
2	$_{	1	,	1	}$	2	&	1	$_{	1	,	0	}$	1	&	6367.036	&	-0.003	&&	3	$_{	1	,	2	}$	3	&	3	$_{	0	,	3	}$	3	&	14807.582	&	0.004	\\
2	$_{	1	,	1	}$	3	&	1	$_{	1	,	0	}$	2	&	6368.254	&	0.005	&&	3	$_{	1	,	2	}$	4	&	3	$_{	0	,	3	}$	3	&	14808.525	&	0.003	\\
2	$_{	1	,	1	}$	1	&	1	$_{	1	,	0	}$	0	&	6369.504	&	0.004	&&	3	$_{	1	,	2	}$	2	&	3	$_{	0	,	3	}$	3	&	14808.855	&	0.002	\\
3	$_{	0	,	3	}$	3	&	2	$_{	0	,	2	}$	3	&	9365.283	&	0.002	&&	4	$_{	1	,	3	}$	4	&	4	$_{	0	,	4	}$	5	&	15055.498	&	0.016	\\
3	$_{	0	,	3	}$	2	&	2	$_{	0	,	2	}$	1	&	9366.345	&	0.004	&&	4	$_{	1	,	3	}$	5	&	4	$_{	0	,	4	}$	5	&	15056.614	&	0.034	\\
3	$_{	0	,	3	}$	3	&	2	$_{	0	,	2	}$	2	&	9366.514	&	-0.021	&&	4	$_{	1	,	3	}$	4	&	4	$_{	0	,	4	}$	4	&	15056.829	&	0.017	\\
3	$_{	0	,	3	}$	4	&	2	$_{	0	,	2	}$	3	&	9366.609	&	0.027	&&	4	$_{	1	,	3	}$	5	&	4	$_{	0	,	4	}$	4	&	15057.899	&	-0.011	\\
3	$_{	0	,	3	}$	2	&	2	$_{	0	,	2	}$	2	&	9368.295	&	0.004	&&	4	$_{	1	,	3	}$	3	&	4	$_{	0	,	4	}$	4	&	15058.201	&	0.008	\\
3	$_{	1	,	3	}$	3	&	2	$_{	1	,	2	}$	3	&	9185.837	&	-0.002	&&	1	$_{	1	,	1	}$	0	&	0	$_{	0	,	0	}$	1	&	17500.635	&	0.000	\\
3	$_{	1	,	3	}$	3	&	2	$_{	1	,	2	}$	2	&	9186.499	&	0.003	&&	1	$_{	1	,	1	}$	2	&	0	$_{	0	,	0	}$	1	&	17501.467	&	-0.003	\\
3	$_{	1	,	3	}$	4,2	&	2	$_{	1	,	2	}$	3,1	&	9186.850	&	0.011	&&	1	$_{	1	,	1	}$	1	&	0	$_{	0	,	0	}$	1	&	17502.021	&	-0.005	\\
3	$_{	1	,	3	}$	2	&	2	$_{	1	,	2	}$	2	&	9187.863	&	0.008	&&	2	$_{	1	,	2	}$	1	&	3	$_{	0	,	3	}$	2	&	4890.715	&	-0.008	\\
3	$_{	1	,	2	}$	3	&	2	$_{	1	,	1	}$	3	&	9550.696	&	0.010	&&	2	$_{	1	,	2	}$	3	&	3	$_{	0	,	3	}$	4	&	4890.805	&	-0.009	\\
3	$_{	1	,	2	}$	3	&	2	$_{	1	,	1	}$	2	&	9551.292	&	0.010	&&	2	$_{	1	,	2	}$	2	&	3	$_{	0	,	3	}$	3	&	4891.448	&	-0.009	\\
3	$_{	1	,	2	}$	2,4	&	2	$_{	1	,	1	}$	1,3	&	9551.640	&	0.011	&&	1	$_{	1	,	1	}$	0	&	2	$_{	0	,	2	}$	1	&	8130.801	&	-0.005	\\
3	$_{	1	,	2	}$	2	&	2	$_{	1	,	1	}$	2	&	9552.569	&	0.013	&&	1	$_{	1	,	1	}$	1	&	2	$_{	0	,	2	}$	1	&	8132.190	&	-0.008	\\
3	$_{	2	,	1	}$	4	&	2	$_{	2	,	0	}$	3	&	9372.956	&	-0.024	&&	1	$_{	1	,	1	}$	2	&	2	$_{	0	,	2	}$	3	&	8132.331	&	-0.006	\\
3	$_{	2	,	2	}$	4	&	2	$_{	2	,	1	}$	3	&	9369.891	&	-0.018	&&	1	$_{	1	,	1	}$	2	&	2	$_{	0	,	2	}$	2	&	8133.586	&	-0.005	\\
4	$_{	0	,	4	}$	4	&	3	$_{	0	,	3	}$	4	&	12483.821	&	0.012	&&	1	$_{	1	,	1	}$	1	&	2	$_{	0	,	2	}$	2	&	8134.142	&	-0.006	\\
4	$_{	0	,	4	}$	5	&	3	$_{	0	,	3	}$	4	&	12485.141	&	0.001	&&	6	$_{	2	,	4	}$	5,7	&	5	$_{	1	,	5	}$	4,6	&	63021.33	&	-0.03	\\
4	$_{	0	,	4	}$	3	&	3	$_{	0	,	3	}$	3	&	12486.785	&	0.003	&&	6	$_{	2	,	4	}$	6	&	5	$_{	1	,	5	}$	5	&	63021.71	&	0.03	\\
4	$_{	1	,	4	}$	4	&	3	$_{	1	,	3	}$	3	&	12247.978	&	-0.002	&&	6	$_{	2	,	5	}$	5,7	&	5	$_{	1	,	4	}$	4,6	&	61143.82	&	-0.02	\\
4	$_{	1	,	4	}$	3	&	3	$_{	1	,	3	}$	2	&	12248.057	&	-0.025	&&	6	$_{	2	,	5	}$	6	&	5	$_{	1	,	4	}$	5	&	61144.06	&	-0.02	\\
4	$_{	1	,	4	}$	5	&	3	$_{	1	,	3	}$	4	&	12248.138	&	0.002	&&	7	$_{	2	,	5	}$	6,8	&	6	$_{	1	,	6	}$	5,7	&	66552.30	&	0.00	\\
4	$_{	1	,	4	}$	3	&	3	$_{	1	,	3	}$	3	&	12249.433	&	-0.009	&&	7	$_{	2	,	5	}$	7	&	6	$_{	1	,	6	}$	6	&	66552.53	&	-0.02	\\
4	$_{	1	,	3	}$	4	&	3	$_{	1	,	2	}$	3	&	12734.359	&	0.015	&&	7	$_{	2	,	6	}$		&	6	$_{	1	,	5	}$		&	63902.50	&	-0.06	\\
4	$_{	1	,	3	}$	3	&	3	$_{	1	,	2	}$	3	&	12735.751	&	0.027	&&	8	$_{	2	,	6	}$	7,9	&	7	$_{	1	,	7	}$	6,8	&	70165.76	&	0.01	\\
4	$_{	2	,	3	}$	5	&	3	$_{	2	,	2	}$	4	&	12492.343	&	-0.020	&&	8	$_{	2	,	6	}$	8	&	7	$_{	1	,	7	}$	7	&	70165.95	&	-0.01	\\
4	$_{	2	,	3	}$	3	&	3	$_{	2	,	2	}$	2	&	12492.473	&	-0.026	&&	8	$_{	2	,	7	}$		&	7	$_{	1	,	6	}$		&	66600.98	&	-0.03	\\
4	$_{	2	,	3	}$	4	&	3	$_{	2	,	2	}$	3	&	12491.798	&	-0.033	&&	9	$_{	2	,	7	}$		&	8	$_{	1	,	8	}$		&	73867.97	&	0.00	\\
4	$_{	2	,	2	}$	4	&	3	$_{	2	,	1	}$	4	&	12499.493	&	-0.016	&&	9	$_{	2	,	8	}$		&	8	$_{	1	,	7	}$		&	69239.74	&	0.06	\\
4	$_{	2	,	2	}$	5	&	3	$_{	2	,	1	}$	4	&	12500.001	&	-0.038	&&	10	$_{	2	,	8	}$		&	9	$_{	1	,	9	}$		&	77665.64	&	-0.04	\\
4	$_{	2	,	2	}$	3	&	3	$_{	2	,	1	}$	2	&	12500.187	&	0.011	&&	10	$_{	2	,	9	}$		&	9	$_{	1	,	8	}$		&	71819.14	&	0.06	\\
5	$_{	0	,	5	}$	6	&	4	$_{	0	,	4	}$	5	&	15600.641	&	0.022	&&	11	$_{	2	,	10	}$		&	10	$_{	1	,	9	}$		&	74339.87	&	0.01	\\
5	$_{	0	,	5	}$	4	&	4	$_{	0	,	4	}$	4	&	15602.195	&	-0.031	&&	12	$_{	2	,	11	}$		&	11	$_{	1	,	10	}$		&	76802.82	&	0.04	\\
5	$_{	1	,	5	}$	4,6	&	4	$_{	1	,	4	}$	3,5	&	15308.669	&	0.009	&&	19	$_{	1	,	18	}$		&	18	$_{	1	,	17	}$		&	60242.00	&	0.03	\\
5	$_{	1	,	4	}$	4,5,6&	4	$_{	1	,	3	}$	3,4,5&	15916.550	&	0.004	&&	19	$_{	2	,	17	}$		&	18	$_{	2	,	16	}$		&	59974.44	&	0.03	\\
1	$_{	1	,	0	}$	1	&	1	$_{	0	,	1	}$	0	&	14498.514	&	0.002	&&	20	$_{	0	,	20	}$		&	19	$_{	0	,	19	}$		&	61565.11	&	-0.02	\\
1	$_{	1	,	0	}$	2	&	1	$_{	0	,	1	}$	2	&	14499.654	&	0.000	&&	20	$_{	1	,	20	}$		&	19	$_{	1	,	19	}$		&	61026.70	&	0.02	\\
1	$_{	1	,	0	}$	0	&	1	$_{	0	,	1	}$	1	&	14499.900	&	-0.004	&&	20	$_{	2	,	19	}$		&	19	$_{	2	,	18	}$		&	62292.35	&	-0.06	\\
\hline
\end{tabular}
\end{footnotesize}
\end{center}
\end{table*}

\begin{table}
\caption[c]{Measured rotational transition frequencies and corresponding deviations from the model of the $^{13}$C1 isotopologue of TC conformer of 4-oxobutanenitrile.}
%\centering
\label{table:rot_lines13C1}
\begin{center}
\begin{footnotesize}
\setlength{\tabcolsep}{10pt}
\begin{tabular}{l l r r }
\hline\hline
$J'_{K'_{\rm a},K'_{\rm c}}$ $F'$&$J"_{K"_{\rm a},K"_{\rm c}}$ $F"$& freq. (MHz) & o.-c. (MHz) \\
\hline
1	$_{	0	,	1	}$	1	&	0	$_{	0	,	0	}$	1	&	3086.854	&	0.007	\\
1	$_{	0	,	1	}$	2	&	0	$_{	0	,	0	}$	1	&	3088.023	&	0.004	\\
1	$_{	0	,	1	}$	0	&	0	$_{	0	,	0	}$	1	&	3089.782	&	0.005	\\
1	$_{	1	,	1	}$	2	&	0	$_{	0	,	0	}$	1	&	17366.338	&	-0.002	\\
1	$_{	1	,	0	}$	1	&	1	$_{	0	,	1	}$	0	&	14397.141	&	-0.004	\\
1	$_{	1	,	0	}$	2	&	1	$_{	0	,	1	}$	2	&	14398.277	&	0.001	\\
1	$_{	1	,	0	}$	0	&	1	$_{	0	,	1	}$	1	&	14398.509	&	0.001	\\
1	$_{	1	,	0	}$	1	&	1	$_{	0	,	1	}$	2	&	14398.895	&	-0.007	\\
1	$_{	1	,	0	}$	2	&	1	$_{	0	,	1	}$	1	&	14399.441	&	-0.006	\\
1	$_{	1	,	0	}$	1	&	1	$_{	0	,	1	}$	1	&	14400.075	&	0.002	\\
2	$_{	0	,	2	}$	2	&	1	$_{	0	,	1	}$	2	&	6173.730	&	0.008	\\
2	$_{	0	,	2	}$	1	&	1	$_{	0	,	1	}$	0	&	6173.925	&	0.007	\\
2	$_{	0	,	2	}$	2	&	1	$_{	0	,	1	}$	1	&	6174.898	&	0.004	\\
2	$_{	0	,	2	}$	3	&	1	$_{	0	,	1	}$	2	&	6174.983	&	0.005	\\
2	$_{	1	,	2	}$	3	&	1	$_{	1	,	1	}$	2	&	6055.928	&	0.009	\\
2	$_{	1	,	1	}$	1	&	2	$_{	0	,	2	}$	1	&	14518.777	&	0.003	\\
2	$_{	1	,	1	}$	3	&	2	$_{	0	,	2	}$	3	&	14519.149	&	0.001	\\
2	$_{	1	,	1	}$	2	&	2	$_{	0	,	2	}$	2	&	14519.830	&	0.011	\\
3	$_{	0	,	3	}$	2	&	2	$_{	0	,	2	}$	1	&	9260.263	&	-0.001	\\
3	$_{	0	,	3	}$	4	&	2	$_{	0	,	2	}$	3	&	9260.519	&	0.013	\\
3	$_{	1	,	3	}$	3	&	2	$_{	1	,	2	}$	2	&	9082.805	&	0.000	\\
3	$_{	1	,	3	}$	2,4	&	2	$_{	1	,	2	}$	1,3	&	9083.175	&	0.027	\\
3	$_{	1	,	2	}$	2	&	2	$_{	1	,	1	}$	1	&	9443.072	&	0.016	\\
4	$_{	0	,	4	}$	3,4,5&	3	$_{	0	,	3	}$	2,3,4&	12343.727	&	-0.023	\\
5	$_{	0	,	5	}$	4,5,6&	4	$_{	0	,	4	}$	3,4,5&	15424.031	&	-0.031	\\
\hline
\end{tabular}
\end{footnotesize}
\end{center}
\end{table}

\begin{table}
\caption[c]{Measured rotational transition frequencies and corresponding deviations from the model of the $^{13}$C2 isotopologue of TC conformer of 4-oxobutanenitrile.}
%\centering
\label{table:rot_lines13C2}
\begin{center}
\begin{footnotesize}
\setlength{\tabcolsep}{10pt}
\begin{tabular}{l l r r }
\hline\hline
$J'_{K'_{\rm a},K'_{\rm c}}$ $F'$&$J"_{K"_{\rm a},K"_{\rm c}}$ $F"$& freq. (MHz) & o.-c. (MHz) \\
\hline
1	$_{	0	,	1	}$	1	&	0	$_{	0	,	0	}$	1	&	3117.900	&	0.002	\\
1	$_{	0	,	1	}$	2	&	0	$_{	0	,	0	}$	1	&	3119.072	&	0.004	\\
1	$_{	0	,	1	}$	0	&	0	$_{	0	,	0	}$	1	&	3120.828	&	0.005	\\
1	$_{	1	,	1	}$	2	&	0	$_{	0	,	0	}$	1	&	17196.779	&	-0.003	\\
1	$_{	1	,	1	}$	1	&	0	$_{	0	,	0	}$	1	&	17197.332	&	-0.007	\\
1	$_{	1	,	0	}$	1	&	1	$_{	0	,	1	}$	0	&	14200.748	&	0.004	\\
1	$_{	1	,	0	}$	2	&	1	$_{	0	,	1	}$	2	&	14201.905	&	0.019	\\
1	$_{	1	,	0	}$	1	&	1	$_{	0	,	1	}$	2	&	14202.489	&	-0.010	\\
2	$_{	0	,	2	}$	2	&	1	$_{	0	,	1	}$	2	&	6235.758	&	0.000	\\
2	$_{	0	,	2	}$	1	&	1	$_{	0	,	1	}$	0	&	6235.954	&	0.001	\\
2	$_{	0	,	2	}$	2	&	1	$_{	0	,	1	}$	1	&	6236.927	&	0.000	\\
2	$_{	0	,	2	}$	3	&	1	$_{	0	,	1	}$	2	&	6237.013	&	0.002	\\
2	$_{	0	,	2	}$	1	&	1	$_{	0	,	1	}$	1	&	6238.874	&	-0.004	\\
2	$_{	1	,	2	}$	3	&	1	$_{	1	,	1	}$	2	&	6113.813	&	0.010	\\
2	$_{	1	,	1	}$	3	&	2	$_{	0	,	2	}$	3	&	14327.034	&	-0.003	\\
2	$_{	1	,	1	}$	2	&	2	$_{	0	,	2	}$	2	&	14327.694	&	0.001	\\
3	$_{	0	,	3	}$	3	&	2	$_{	0	,	2	}$	3	&	9352.083	&	-0.010	\\
3	$_{	0	,	3	}$	2	&	2	$_{	0	,	2	}$	1	&	9353.144	&	-0.008	\\
3	$_{	0	,	3	}$	4	&	2	$_{	0	,	2	}$	3	&	9353.403	&	0.009	\\
3	$_{	0	,	3	}$	2	&	2	$_{	0	,	2	}$	3	&	9353.844	&	-0.005	\\
3	$_{	1	,	3	}$	2,4	&	2	$_{	1	,	2	}$	1,3	&	9169.938	&	0.0024	\\
3	$_{	1	,	2	}$	2,4	&	2	$_{	1	,	1	}$	1,3	&	9542.487	&	0.0070	\\
4	$_{	0	,	4	}$	3,4,5&	3	$_{	0	,	3	}$	2,3,4	&	12467.297	&	0.0008	\\
\hline
\end{tabular}
\end{footnotesize}
\end{center}
\end{table}

\begin{table}
\caption[c]{Measured rotational transition frequencies and corresponding deviations from the model of the $^{13}$C3 isotopologue of TC conformer of 4-oxobutanenitrile.}
%\centering
\label{table:rot_lines13C3}
\begin{center}
\begin{footnotesize}
\setlength{\tabcolsep}{10pt}
\begin{tabular}{l l r r }
\hline\hline
$J'_{K'_{\rm a},K'_{\rm c}}$ $F'$&$J"_{K"_{\rm a},K"_{\rm c}}$ $F"$& freq. (MHz) & o.-c. (MHz) \\
\hline
1	$_{	0	,	1	}$	1	&	0	$_{	0	,	0	}$	1	&	3119.657	&	-0.001	\\
1	$_{	0	,	1	}$	2	&	0	$_{	0	,	0	}$	1	&	3120.828	&	0.001	\\
1	$_{	0	,	1	}$	0	&	0	$_{	0	,	0	}$	1	&	3122.577	&	-0.003	\\
1	$_{	1	,	1	}$	1	&	0	$_{	0	,	0	}$	1	&	17377.215	&	0.004	\\
1	$_{	1	,	0	}$	1	&	1	$_{	0	,	1	}$	0	&	14377.229	&	-0.004	\\
1	$_{	1	,	0	}$	2	&	1	$_{	0	,	1	}$	2	&	14378.369	&	-0.003	\\
1	$_{	1	,	0	}$	1	&	1	$_{	0	,	1	}$	2	&	14378.993	&	0.007	\\
1	$_{	1	,	0	}$	2	&	1	$_{	0	,	1	}$	1	&	14379.530	&	-0.011	\\
1	$_{	1	,	0	}$	1	&	1	$_{	0	,	1	}$	1	&	14380.154	&	-0.001	\\
2	$_{	0	,	2	}$	2	&	1	$_{	0	,	1	}$	2	&	6239.302	&	-0.007	\\
2	$_{	0	,	2	}$	1	&	1	$_{	0	,	1	}$	0	&	6239.507	&	0.004	\\
2	$_{	0	,	2	}$	2	&	1	$_{	0	,	1	}$	1	&	6240.477	&	0.000	\\
2	$_{	0	,	2	}$	3	&	1	$_{	0	,	1	}$	2	&	6240.563	&	0.002	\\
2	$_{	0	,	2	}$	1	&	1	$_{	0	,	1	}$	1	&	6242.428	&	0.003	\\
2	$_{	1	,	2	}$	2	&	1	$_{	1	,	1	}$	1	&	6117.760	&	0.022	\\
2	$_{	1	,	2	}$	3	&	1	$_{	1	,	1	}$	2	&	6118.949	&	-0.001	\\
2	$_{	1	,	1	}$	3	&	2	$_{	0	,	2	}$	3	&	14501.870	&	0.008	\\
3	$_{	0	,	3	}$	3	&	2	$_{	0	,	2	}$	3	&	9357.485	&	-0.012	\\
3	$_{	0	,	3	}$	2	&	2	$_{	0	,	2	}$	1	&	9358.540	&	-0.015	\\
3	$_{	0	,	3	}$	4	&	2	$_{	0	,	2	}$	3	&	9358.803	&	0.007	\\
3	$_{	1	,	3	}$	3	&	2	$_{	1	,	2	}$	2	&	9177.338	&	0.004	\\
3	$_{	1	,	3	}$	2,4	&	2	$_{	1	,	2	}$	1,3	&	9177.669	&	-0.007	\\
3	$_{	1	,	2	}$	2,4	&	2	$_{	1	,	1	}$	1,3	&	9545.342	&	0.008	\\
\hline
\end{tabular}
\end{footnotesize}
\end{center}
\end{table}

\begin{table}
\caption[c]{Measured rotational transition frequencies and corresponding deviations from the model of the $^{13}$C4 isotopologue of TC conformer of 4-oxobutanenitrile.}
%\centering
\label{table:rot_lines13C4}
\begin{center}
\begin{footnotesize}
\setlength{\tabcolsep}{10pt}
\begin{tabular}{l l r r }
\hline\hline
$J'_{K'_{\rm a},K'_{\rm c}}$ $F'$&$J"_{K"_{\rm a},K"_{\rm c}}$ $F"$& freq. (MHz) & o.-c. (MHz) \\
\hline
1	$_{	0	,	1	}$	1	&	0	$_{	0	,	0	}$	1	&	3089.956	&	0.005	\\
1	$_{	0	,	1	}$	2	&	0	$_{	0	,	0	}$	1	&	3091.122	&	0.004	\\
1	$_{	0	,	1	}$	0	&	0	$_{	0	,	0	}$	1	&	3092.870	&	-0.001	\\
1	$_{	1	,	1	}$	2	&	0	$_{	0	,	0	}$	1	&	17467.348	&	-0.001	\\
1	$_{	1	,	1	}$	1	&	0	$_{	0	,	0	}$	1	&	17467.913	&	0.012	\\
1	$_{	1	,	0	}$	1	&	1	$_{	0	,	1	}$	0	&	14494.372	&	0.006	\\
1	$_{	1	,	0	}$	2	&	1	$_{	0	,	1	}$	2	&	14495.506	&	0.004	\\
1	$_{	1	,	0	}$	0	&	1	$_{	0	,	1	}$	1	&	14495.742	&	-0.006	\\
1	$_{	1	,	0	}$	1	&	1	$_{	0	,	1	}$	2	&	14496.110	&	-0.008	\\
2	$_{	0	,	2	}$	1	&	1	$_{	0	,	1	}$	0	&	6180.140	&	0.005	\\
2	$_{	0	,	2	}$	2	&	1	$_{	0	,	1	}$	1	&	6181.116	&	0.008	\\
2	$_{	0	,	2	}$	3	&	1	$_{	0	,	1	}$	2	&	6181.199	&	0.007	\\
2	$_{	0	,	2	}$	1	&	1	$_{	0	,	1	}$	1	&	6183.063	&	0.008	\\
2	$_{	1	,	1	}$	1	&	2	$_{	0	,	2	}$	1	&	14615.309	&	-0.001	\\
2	$_{	1	,	1	}$	3	&	2	$_{	0	,	2	}$	3	&	14615.671	&	-0.005	\\
2	$_{	1	,	1	}$	2	&	2	$_{	0	,	2	}$	2	&	14616.335	&	0.000	\\
3	$_{	0	,	3	}$	2	&	2	$_{	0	,	2	}$	1	&	9269.619	&	-0.001	\\
3	$_{	0	,	3	}$	3	&	2	$_{	0	,	2	}$	2	&	9269.802	&	-0.012	\\
3	$_{	0	,	3	}$	4	&	2	$_{	0	,	2	}$	3	&	9269.878	&	0.017	\\
3	$_{	1	,	2	}$	3	&	2	$_{	1	,	1	}$	2	&	9450.997	&	0.008	\\
3	$_{	1	,	2	}$	2,4	&	2	$_{	1	,	1	}$	1,3	&	9451.335	&	0.000	\\
4	$_{	0	,	4	}$	3	&	3	$_{	0	,	3	}$	2	&	12356.203	&	-0.018	\\
4	$_{	0	,	4	}$	5	&	3	$_{	0	,	3	}$	4	&	12356.324	&	-0.010	\\
\hline
\end{tabular}
\end{footnotesize}
\end{center}
\end{table}

\begin{table}
\caption[c]{Measured rotational transition frequencies and corresponding deviations from the model of the $^{18}$O isotopologue of TC conformer of 4-oxobutanenitrile.}
%\centering
\label{table:rot_lines18O}
\begin{center}
\begin{footnotesize}
\setlength{\tabcolsep}{10pt}
\begin{tabular}{l l r r }
\hline\hline
$J'_{K'_{\rm a},K'_{\rm c}}$ $F'$&$J"_{K"_{\rm a},K"_{\rm c}}$ $F"$& freq. (MHz) & o.-c. (MHz) \\
\hline
1	$_{	1	,	1	}$	2	&	0	$_{	0	,	0	}$	1	&	17087.181	&	0.000	\\
2	$_{	0	,	2	}$	1	&	1	$_{	0	,	1	}$	0	&	6032.992	&	0.003	\\
2	$_{	0	,	2	}$	2	&	1	$_{	0	,	1	}$	1	&	6033.979	&	0.004	\\
2	$_{	0	,	2	}$	3	&	1	$_{	0	,	1	}$	2	&	6034.071	&	0.011	\\
2	$_{	0	,	2	}$	1	&	1	$_{	0	,	1	}$	1	&	6035.949	&	0.001	\\
3	$_{	0	,	3	}$	4	&	2	$_{	0	,	2	}$	3	&	9049.217	&	-0.013	\\
3	$_{	1	,	3	}$	2,4	&	2	$_{	1	,	2	}$	1,3	&	8878.004	&	0.000	\\
\hline
\end{tabular}
\end{footnotesize}
\end{center}
\end{table}

\begin{table}
\caption[c]{Measured rotational transition frequencies and corresponding deviations from the model of the $^{15}$N isotopologue of TC conformer of 4-oxobutanenitrile.}
%\centering
\label{table:rot_lines15N}
\begin{center}
\begin{footnotesize}
\setlength{\tabcolsep}{10pt}
\begin{tabular}{l l r r }
\hline\hline
$J'_{K'_{\rm a},K'_{\rm c}}$&$J"_{K"_{\rm a},K"_{\rm c}}$ & freq. (MHz) & o.-c. (MHz) \\
\hline
1	$_{	0	,	1	}$		&	0	$_{	0	,	0	}$		&	3041.633	&	0.003	\\
1	$_{	1	,	1	}$		&	0	$_{	0	,	0	}$		&	17461.238	&	0.018	\\
1	$_{	1	,	0	}$		&	1	$_{	0	,	1	}$		&	14534.928	&	-0.008	\\
2	$_{	0	,	2	}$		&	1	$_{	0	,	1	}$		&	6082.579	&	0.008	\\
2	$_{	1	,	1	}$		&	2	$_{	0	,	2	}$		&	14650.961	&	-0.010	\\
3	$_{	0	,	3	}$		&	2	$_{	0	,	2	}$		&	9122.134	&	0.001	\\
3	$_{	1	,	2	}$		&	2	$_{	1	,	1	}$		&	9297.493	&	0.017	\\
4	$_{	0	,	4	}$		&	3	$_{	0	,	3	}$		&	12159.606	&	-0.023	\\
\hline
\end{tabular}
\end{footnotesize}
\end{center}
\end{table}

\begin{table}
\caption[c]{Experimental spectroscopic parameters of the $^{13}$C, $^{15}$N, and $^{18}$O isotopologues of the TC conformer of 4-oxobutanenitrile, derived from the fit of its rotational transitions recorded in the 2$-$18 GHz range.}
%\centering
\label{table:spectroscopic_iso}
\begin{center}
\begin{footnotesize}
\setlength{\tabcolsep}{1pt}
\begin{tabular}{l r r r}
\hline\hline
Parameter & $^{13}$C1 & $^{13}$C2 & $^{13}$C3 \\
\hline
$A_{\rm 0}$  (MHz)                               &   15882.5029(43) &  15699.5295(27)  &  15877.7092(32)    \\
$B_{\rm 0}$ (MHz)                                &   1603.8960(20)  &   1621.5271(10)  &  1621.5922(15)     \\
$C_{\rm 0}$  (MHz)                               &   1483.9275(17)  &   1497.3456(10)  &  1499.0399(12)     \\
$\frac{3}{2}\chi_{\rm aa}$ (MHz)                 &      -5.859(19)  &      -5.8499(98) &     -5.843(11)     \\
$\frac{1}{4}(\chi_{\rm bb}-\chi_{\rm cc})$ (MHz) &      -0.0676(75) &      -0.0461(70) &     -0.0504(79)    \\
$N$                                              &      25          &    24            &     23             \\
$\sigma$  (kHz)                                  &      11          &    7             &      8             \\
\hline
Parameter & $^{13}$C4 & $^{18}$O & $^{15}$N \\
\hline
$A_{\rm 0}$  (MHz)                               &    15981.6204(32) &  15636.5033(73)  &   15998.0781(83)    \\
$B_{\rm 0}$ (MHz)                                &     1605.1034(13) &   1566.5668(32)  &    1578.4878(32)    \\
$C_{\rm 0}$  (MHz)                               &     1485.8203(15) &   1450.7766(24)  &    1463.1422(36)    \\
$\frac{3}{2}\chi_{\rm aa}$ (MHz)                 &       -5.840(13)  &     -5.918(19)   &     ...            \\
$\frac{1}{4}(\chi_{\rm bb}-\chi_{\rm cc})$ (MHz) &       -0.0520(54) &     [0]$^a$      &     ...            \\
$N$                                              &       23          &      7           &      8             \\
$\sigma$  (kHz)                                  &        8          &      7           &     13             \\
\hline
\end{tabular}
\end{footnotesize}
\end{center}
$^a$ Fixed to zero in the fit.
\end{table}

%If you want to present additional material which would interrupt the flow of the main paper, it can be placed in an Appendix which appears after the list of references.

%%%%%%%%%%%%%%%%%%%%%%%%%%%%%%%%%%%%%%%%%%%%%%%%%%

% Don't change these lines
\bsp	% typesetting comment
\label{lastpage}
\end{document}